# Can a Matter-Wave Interferometer Detect Translational Speed?

Ruyong Wang*, Yi Zheng and Aiping Yao
*St. Cloud State University, St. Cloud, MN 56301, USA*

Based only on the Galilean addition of velocities and the de Broglie relation, it is deduced that in a matter-wave interferometer with slow-speed particles, a moving segment of $\Delta L$ with a velocity $V$ contributes $\Delta\phi = (2\pi/v\lambda)V \cdot \Delta L$ to the total phase difference of the interferometer, where $v$ is the speed of the particles and $\lambda$ is the wavelength. This expression is exactly the same as the generalized Sagnac effect for light waves found by experiments except that $v$ is replaced by $c$. For a rotational motion, it leads to the Sagnac effect. Additionally, the scientific value of this relationship is also to explore the possibility of detecting translation speeds by a matter-wave interferometer. Two configurations of the experimental setup have been indicated and the key element is that the paths of the interfering beams constitute a loop with an opening. If the possibility is confirmed by experiments, the conclusions will be that there is a preferred reference frame for matter waves and a speedometer with a very high sensitivity is possible.

That a matter-wave interferometer can detect the rotation of the Earth was first predicted in 1975 and the formula of the phase shift was obtained by an analogue of the Sagnac effect for light waves [1]. The prediction was confirmed by an experiment using a neutron interferometer in 1979 [2]. Since then the Sagnac effect of matter waves has also been shown in the atom interferometer [3] and the electron interferometer [4]. Similar to the Sagnac effect of light waves which has a variety of interpretations, the Sagnac effect of matter waves also has different interpretations, some of which are common for both light waves and matter waves [4, 5]. Here we present a concise interpretation based directly on only two fundamental equations, the Galilean addition of velocities and the de Broglie relations. Because the Sagnac experiments of matter waves use slow-speed particles, the relativistic effects do not need to be considered.

For a matter wave in a segment of a matter-wave interferometer that is at rest in an inertial frame $\Sigma$ (Fig. 1), we have the de Broglie wavelength:

$$\lambda = h/mv \qquad (1)$$

where $m$ is the mass of a particle and $v$ is the speed of particles. The phase shift in the segment of the length $\Delta L$ is

$$\phi = 2\pi(\Delta L/\lambda). \tag{2}$$

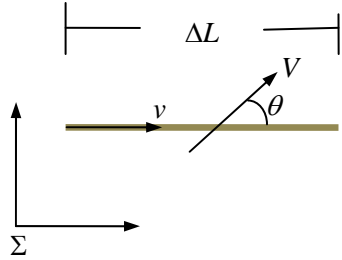

Fig. 1 A moving segment of the matter-wave interferometer.

Now allow this segment of the interferometer to move with velocity $V$ in the reference frame $\Sigma$. According to the Galilean addition of velocities, the speed of the particles in the direction of the matter wave is $v+V\cos\theta$. Therefore we have a new wavelength

$$\lambda' = h/m(v+V\cos\theta) = \lambda/[1+(V\cos\theta)/v] \tag{3}$$

Because the total energy of the particle is almost the same, the frequency of the matter wave does not change. Thus, the phase shift in the moving segment is

$$\phi' = 2\pi(\Delta L/\lambda') = 2\pi(\Delta L/\lambda)[1+(V\cos\theta)/v] \tag{4}$$

Hence for the segment $\Delta L$, the phase difference between the two cases is

$$\Delta\phi = \phi' - \phi = 2\pi\,[(V\cos\theta)/v](\Delta L/\lambda) = (2\pi/v\lambda)\mathbf{V}\cdot\Delta\mathbf{L} \tag{5}$$

Here, the magnitude of vector $\Delta\mathbf{L}$ is the length of the segment, and its direction is along the propagating matter wave. This expression tells us that when the direction of the matter wave changes to the opposite direction, i.e., $\Delta\mathbf{L}$ changes to $-\Delta\mathbf{L}$, the phase difference changes its sign.

Note that this expression is exactly the same as the expression of the generalized Sagnac effect (GSE) obtained by experiments for light waves [6, 7] except that $v$ is replaced by $c$. The GSE states that in a loop a moving segment of $\Delta\mathbf{L}$ with a velocity $\mathbf{V}$ contributes $\Delta\phi = (4\pi/c\lambda)\mathbf{V}\cdot\Delta\mathbf{L}$ to the total phase difference between two counter-propagating beams in the loop (the phase difference is doubled because of two counter-propagating beams in the interferometer). The contribution $\Delta\phi$ is independent of the refractive index of the waveguide. Furthermore the motion of the segment can be either linear or circular. The GSE includes the Sagnac effect of rotation as a special case.

Similarly, for the rotation of the matter-wave interferometer (Fig. 2), the phase difference between the two paths is

$$\Delta\phi = \frac{2\pi}{v\lambda}\left(\int_{ABC} \boldsymbol{V}\cdot d\boldsymbol{l} - \int_{ADC} \boldsymbol{V}\cdot d\boldsymbol{l}\right) = \frac{2\pi}{v\lambda}\left(\int_{ABC} \boldsymbol{V}\cdot d\boldsymbol{l} + \int_{CDA} \boldsymbol{V}\cdot d\boldsymbol{l}\right) = \frac{2\pi}{v\lambda}\oint \boldsymbol{V}\cdot d\boldsymbol{l} \qquad (6)$$

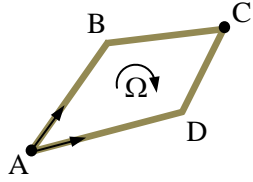

Fig. 2 The Sagnac experiment of the matter-wave interferometer.

Applying Stokes's theorem, we obtain the Sagnac effect of rotation for matter waves

$$\Delta\phi = \frac{2\pi}{v\lambda}\oint \boldsymbol{V}\cdot d\boldsymbol{l} = \frac{2\pi}{v\lambda}\iint_A (\nabla\times\boldsymbol{V})\cdot d\boldsymbol{A} = \frac{2\pi}{v\lambda}\iint_A 2\boldsymbol{\Omega}\cdot d\boldsymbol{A} = \frac{4\pi}{v\lambda}\boldsymbol{\Omega}\cdot\boldsymbol{A} = \frac{2m}{\hbar}\boldsymbol{\Omega}\cdot\boldsymbol{A} \qquad (7)$$

where $\boldsymbol{\Omega}$ is the angular velocity of the rotation and $A$ is the area enclosed by the paths of the interferometer.

While the enclosed area is widely cited as a factor of the Sagnac effect, experiments for the GSE indicate that the velocity and length of the moving waveguide are the fundamental factors, rather than the enclosed area [6]. Actually, the experiments show that the enclosed area can be zero for a non-zero phase shift caused by the motion of the interferometer's segments. Similarly, this work shows that the fundamental factors for the Sagnac effect of matter waves are not the speed of the rotation and the enclosed area, but the velocity and the length of the path. The velocity changes the momentum of the particles, which causes the change of the wavelength of matter waves, which in turn causes the phase difference in the interferometer.

The scientific merits of this concise and direct analysis are not only in interpreting the Sagnac effect of rotation, but also in exploring the possibility of detecting translation speeds by a matter-wave interferometer, i.e., the possibility of existence of a preferred reference frame for matter waves. It can be shown that if the matter-wave interferometer is a Mach-Zehnder type interferometer with paths constituting a closed loop (Fig. 3a), a matter-wave interferometer cannot detect translational speeds. The phase difference between two paths from A to B is

$$\Delta\phi = \Delta\phi_{II} - \Delta\phi_I = \frac{2\pi}{v\lambda}\left(\int_{II} \mathbf{V}\cdot d\mathbf{l} - \int_I \mathbf{V}\cdot d\mathbf{l}\right) \tag{8}$$

Because in the case of a translational motion all the segments have the same velocity $\mathbf{V}$, we have

$$\Delta\phi = \frac{2\pi}{v\lambda}\mathbf{V}\cdot\left(\int_{II} d\mathbf{l} - \int_I d\mathbf{l}\right) = 0 \tag{9}$$

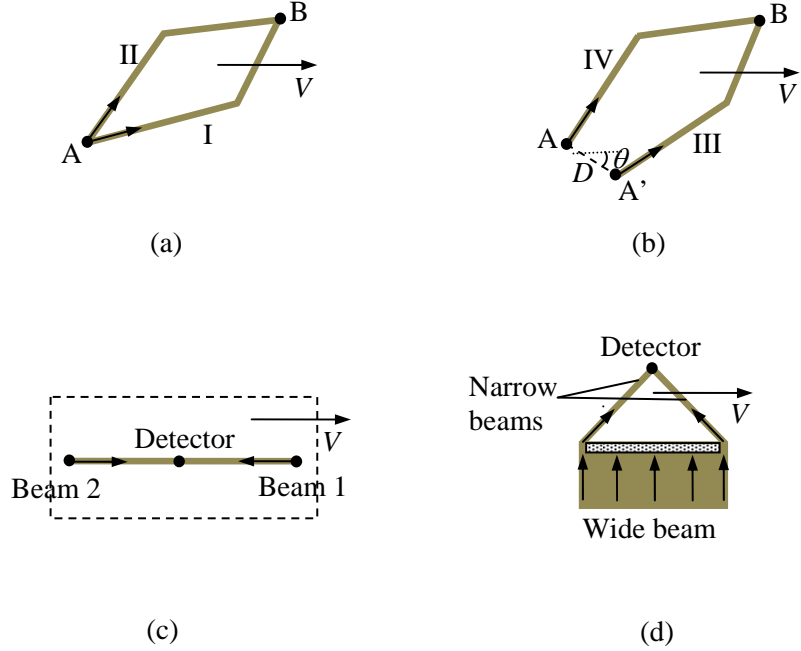

(a) (b) (c) (d)

Fig. 3 Detecting the translational motion with a matter-wave interferometer. (a) A Mach-Zehnder type interferometer without an opening and (b) with an opening. (c) Interference between two independent beams. (d) Coherently extracting two separated "narrow" beams from a "wide" beam.

However, if the paths of the beams constitute a loop with an opening, for example, with a distance $D$ between the starting points of the two beams (Fig. 3b), the translational velocity $\mathbf{V}$ will have a net effect for the interferometer:

$$\Delta\phi' = \frac{2\pi}{v\lambda}\mathbf{V}\cdot\left(\int_{IV} d\mathbf{l} - \int_{III} d\mathbf{l}\right) = \frac{2\pi}{v\lambda}VD\cos\theta = \left(\frac{m}{\hbar}\right)VD\cos\theta \tag{10}$$

The possible configurations having two beams with an initial distance *D* include: (1) utilizing the interference between two independent beams [8], especially two separated atom lasers which interfere with each other [9] as shown in Fig. 3c, and (2) coherently extracting two separated "narrow" beams from a "wide" beam as shown in Fig. 3d.

A typical value for $v\lambda$ of matter waves is $10^{-8}$ m$^2$/s, therefore even if the size of the opening *D* is only 100 μm, experiments should be able to examine whether $\Delta\phi = (2\pi/v\lambda)VD\cos\theta$ exists or not with a speed *V* slower than $10^{-2}$ cm/s. And if this effect is confirmed by experiments, using the matter-wave interferometer we could have a speedometer with a very high sensitivity.

It would seem that detecting translational speeds using a matter-wave interferometer having slow-speed particles is impossible because it violates the classical principle of relativity. However, we know that the classical principle of relativity is valid for the phenomena of mechanics; i.e., it is based on the particle–like property of matter. It has not been examined by experiments whether or not the classical principle of relativity is valid for matter waves. Fig. 3c and 3d are the proposed experiments for the examining. If the fringe shifts are found by the experiment, the conclusion will be that there is a preferred reference frame for matter waves.

*Electronic address: ruwang@stcloudstate.edu**References**

[1] L. A. Page, *Phys. Rev. Lett.* **35**, 543 (1975).
[2] S. A. Werner, J.-L. Staudenmann, and R. Colella, *Phys. Rev. Lett.* **42**, 1103 (1979).
[3] F. Riehle, Th. Kisters, A. Witte, J. Helmcke, and Ch. J. Bordé, *Phys. Rev. Lett.* **67**, 177 (1991).
[4] F. Hasselbach and M. Nicklaus, *Phys. Rev.* A **48**, 143 (1993).
[5] S. Werner, *Gen. Relativ. Gravit.* **40**, 921 (2008).
[6] R. Wang, Y. Zheng, A. Yao, D. Langley, *Phys. Lett. A* **312**, 7 (2003).
[7] R. Wang, Y. Zheng, A. Yao, *Phys. Rev. Lett.* **93**, 143901 (2004).
[8] M. R. Andrews, C. G. Townsend, H.-J. Miesner, D. S. Durfee, D. M. Kurn, and W. Ketterle, *Science* **275**, 637 (1997).
[9] W. Ketterle, private e-mail communication, Aug. 2008.5